\baselineskip=2\baselineskip
\magnification 1200
\rightline{RU-95-59}
\vskip 0.4in
\centerline{\bf Boundary S-matrix for the Tricritical Ising Model}
\vskip 0.6in
\centerline{Leung Chim \footnote{$^\dagger$}{E-mail:
CHIM@ruhets.rutgers.edu}}
\vskip 0.1in
\centerline{Department of Physics and Astronomy}
\centerline{Rutgers University}
\centerline{P.O.Box 849, Piscataway, NJ 08855-0849}
\vskip 0.4in
\def\b{\bar}
\def\u{\upsilon}
\def\h{\theta}
\def\r{\rangle}
\vskip 0.1in
\centerline{\bf Abstract}

The Tricritical Ising model perturbed by the subleading energy operator
$\Phi_{({3\over5})}$ was known to be
an Integrable Scattering Theory of massive kinks [14] and in fact
preserves supersymmetry. We consider here the model defined on the
half-plane with a boundary and computed the associated factorizable
boundary S-matrix. The conformal boundary conditions of this model
were identified and the corresponding S-matrices were found.
We also show how some of these S-matrices can be perturbed
and generate ``flows'' between different boundary conditions.

\hbox{}\hfil\break

{\bf 1. Introduction}
\hbox{}\hfil\break

Quantum field theory of systems with boundaries have recieved much
attention in recent years due to substantial advances in several
apsects of this subject. One successful area
of research is in the study of
s-wave scattering of electrons from impurities [1] employing the
technology of boundary conformal field theory [2]. Another
fruitful avenue of work is the calculation of the boundary S-matrix
for the elastic reflection of particles off the boundary [3]. In the
context of quantum spin chains, this matrix is known as the $K$ matrix
[4, 5]. Such $K$ matrix facuilties the generalization of the quantum
inverse scattering method to the case of an open spin chain [6]-[9].
While for the case of integrable perturbation of conformal field theory,
the boundary S-matrix is factorizable if integrable boundary conditions
are choosen [10]. By applying the technique of thermodynamic Bethe
ansatz with the boundary S-matirx, many physical results relating to
boundary phenomena can be obtained [11]-[13].

The current work concerns the boundary S-matrix for the tricritical
Ising model. As is known, perturbing this model in the bulk
by the sub-leading energy operator ${\u'}$ generates a flow from
the $c = {{7\over{10}}}$ conformal field theory down to a
field theory of massive kinks [14].
\footnote{$^1$}{The same perturbation with an
opposite sign in the coupling constant generates a massless flow down to
the Ising fixed point [15, 16]}
In fact the resultant theory is supersymmertic and the supercharges can
be explicitly constructed as integrals of motion [14].
In the present work, we consider
this model defined on the half-plane with a boundary.
The general S-matrix
describing the factorizable scattering of kinks from the wall is derived
by solving the corresponding ``boundary Yang-Baxter equation''.
The conformal boundary conditions of the model can be classified following
the proposal of [2]. This is done by using the RSOS picture of the Tricritical
Ising model. The S-matrix associated with the conformal boundary conditions
were found and we also studied the perturbation of these boundary conditions
by a relevant boundary field.

\hbox{}\hfil\break
{\bf 2. Backround on the Tricritical Ising Model}

\hbox{}\hfil\break
The 2-Dimensional Tricritical Ising Model is of interest to physicists
because it describes tricritical phenomena in a variety of microscopic
models [17].
In particular, it is the simplest known statistical model to exhibit
Supersymmetry [18], and the fact that it can be realized experimentally
[19] makes it an important model to study.

Several lattice realization of this tricritical behavior is possible,
among them the Blume-Capel Quantum Spin Chain [20] and the $A_4$
Restricted Solid-on-Solid (RSOS) model [21]. The microscopic model
that we will consider here is the Ising model with annealed vacancies
[22]. The classical Hamiltonian is
$$H = -\beta \sum_{<i,j>}\sigma_i\sigma_j - \mu \sum_{i}(\sigma_i)^2
, \eqno(1)$$
where $i,j$ label the lattice sites, each of the variables $\sigma_i$
takes three values $-1,0,1$ ($0$ represents the vacant site) and the
first sum in (1) is made over nearest neighbor pairs. Concentration
of the vacancies is controlled by the chemical potential $\mu$. The
phase diagram of this system in coordinates of temperature $T$ and
$e^{-\mu}$ is shown (schematically) in Fig.1. There are two phases
separated by a line AB of phase transition. For low vacancies and low
temperature, we have an ordered phase I with spontaneously broken $Z_2$
symmetry. On the other side of the transition line we have a disordered
phase II with unbroken symmetry. The point B is the usual critical point
of the Ising model, and the solid segment BT corresponds to the second
order transition belonging to the Ising model universality class. The
dotted segment AT is a line of first order phase transition. The point T
where the two critical segments joined smoothly is the tricritical point of the
system. This model belongs to the universality class of the
Landau-Ginzburg $\varphi^6$-theory at this tricritical point [23], so
the phase diagram could also be understood in terms
of the Landau-Ginzburg effective potential of this system [14].
This effective potential (specific free energy) is plotted
as the function of the order parameter ($<\sigma>$) of this system in Fig.2
for various regions of the phase diagram.
The ground state is degenerate in phase I,
while in phase II it is non-degenerate. There is only one vacuum
on the line BT, typical of a second order transition.
Note that along the first
order transition line AT, the vacuum is in fact three-fold degenerate.

At the tricritical point T, the system is described by the conformal field
theory (CFT) with central charge $c={7\over10}$ [24]. There are six irreducible
representations of the Virasoro algebra, and the corresponding conformal
dimensions $\Delta_{(r,s)}$ of the primary field $\phi_{(r,s)}$ are organized
into a Kac table in Fig.3.
\vskip 0.2 in

\hskip 1.5 in\vbox{
\tabskip=0pt \offinterlineskip
\def\hr{\noalign{\hrule}}
\def\o{\over}
\halign
       {\vrule# & \hskip1em\hfil#\hskip1em&
        \vrule# & \hskip1em\hfil#\hskip1em&
        \vrule# & \hskip1em\hfil#\hskip1em&
        \vrule# & \hskip1em\hfil#\hskip1em&
        \vrule#\cr
        \hr
        height 14 pt &${3\o2}$ && ${3\o5}$ && ${1\o10}$ && $0$ &\cr
        height 2 pt &\omit && \omit && \omit && \omit &\cr
        \hr
        height 14 pt &${7\o16}$ && ${3\o80}$ && ${3\o80}$ && ${7\o16}$ &\cr
        height 2 pt &\omit && \omit && \omit && \omit &\cr
        \hr
        height 14 pt &${0}$ && ${1\o10}$ && ${3\o5}$ && ${3\o2}$ &\cr
        height 2 pt &\omit && \omit && \omit && \omit &\cr
        \hr}
}

\centerline{{\bf Figure} 3}

The fields in this theory can be classified
according to their properties under the $Z_2$ spin-reversal transformation
$\sigma \to -\sigma$ (this is a symmetry of the microscopic model (1)).
In particular we have four even fields:
the identity $I \equiv \Phi_{(0)}$, the leading
energy density $\epsilon \equiv \Phi_{({1\over10})}$, the subleading
energy (vacancy) density $\epsilon' \equiv \Phi_{({3\over5})}$
and the irrelevant field $\epsilon'' \equiv \Phi_{({3\over2})}$.
Here we use the short-hand notation $\Phi_{(\Delta)}$ for the scalar field
$\Phi_{(\Delta,\b\Delta)}$ with $\Delta=\b\Delta$.
There are two odd fields under spin-reversal: the leading spin field
$\sigma \equiv \Phi_{({3\over80})}$, and the subleading spin field $\sigma'
\equiv \Phi_{({7\over16})}$.

In addition to the scalar fields, this conformal
field theory also possesses
fermion fields $G(z) \equiv \Phi_{({3\over2},0)}$ and $\bar{G}(\bar{z}) \equiv
\Phi_{(0,{3\over2})}$. The currents $G(z)$ ($\bar{G}(\bar{z})$)
together with the stress
tensor $T(z)$ ($\bar{T}(\bar{z})$) generate the Neveu-Schwartz-Ramond algebra,
hence this model exhibits superconformal symmetry [18]. We can alternatively
classify the fields in this CFT according to this extended symmetry. For
example, in the Neveu-Schwartz sector of this CFT we have the superfield
$$\bf\Phi_{({1\over10})}(z,\h;\bar{z},\bar{\h}) =
\epsilon + \h\b\psi + \bar{\theta}\psi + \theta\bar{\theta}\epsilon',$$
where $\psi \equiv \Phi_{({3\over5},{1\over10})}$, and $\bar\psi \equiv
\Phi_{({1\over10},{3\over5})}$ are fermion and antifermion fields respectively.
All the component fields of $\bf\Phi_{({1\over10})}$ are mutually local and,
together with the identity operator $I$ and its descendants, they constitute
the Neveu-Schwartz sector of this CFT. In this sector, Kramers-Wannier duality
transformation
acts as a second $Z_2$ symmetry under which $\bf\Phi_{({1\over10})} \to
-\bf\Phi_{({1\over10})}$; $\theta \to -\theta$. More specifically, under
duality
transformation, the signs of $\epsilon, \Psi$, and $\bar{G}$ are changed, while
$I, \epsilon', \bar\Psi$, and $G$ are unaltered.

The spin fields $\sigma$ and $\sigma'$ provide the Ramond representation
of the superconformal symmetry [18]. The duality transformation converts them
into the disorder fields $\mu$ and $\mu'$ respectively [25], where
$\mu = G_{0}\sigma$, $\mu' = G_0\sigma'$. Here $G_0$ is the zero-mode
of $G$ in the Ramond sector. Hence the Operator Product Algebras generated by
$(I, \sigma, \sigma', \epsilon, \epsilon')$ and
$(I, \mu, \mu', -\epsilon, \epsilon')$ are isomorphic under duality.
In other words, the Tricritical Ising Model is a self-dual system.
\hbox{}\hfil\break

{\bf 3. Supersymmetric perturbation of the Tricritical Ising Model}
\hbox{}\hfil\break

In the ${c={7\over10}}$ conformal field theory,
the energy densities and spin fields
are relevant operators. Perturbation of the bulk theory by these operators
were studied numerically in [26].
This work confirms the conjecture that perturbations by the operators
$\epsilon$, $\epsilon'$ and $\sigma'$ are individually integrable [27].
In particular, the transition line AB in Fig.1 can be generated by perturbing
the tricritical fixed point Hamiltonian by the vacancy density $\epsilon'$
[15]. Thus the scaling region around the point T can be described by the
field theory with the ``action''
$$H_{\lambda} = H_{({7\over10})} + \lambda \int \Phi_{({3\over5})} {d^2}x,
\eqno(2)$$
where $H_{({7\over10})}$ is the ``action'' of the bulk CFT. The renormalization
group (RG) flow from T down to the Ising fixed point B is given by (2) with
$\lambda>0$ and can be described by a massless scattering field theory [16].
For $\lambda<0$, (2) is an integrable massive field theory [14] and is
associated with the first order transition line AT. According to [27], this
model possesses an infinite number of bosonic Integrals of Motion, and we shall
concentrate on the case with $\lambda<0$. Further more, because
$\Phi_{({3\over5})}$ is the upper component of the superfield
$\bf{\Phi}_{({1\over10})}$,  so the
perturbation (5.2) does not destroy the global supersymmetry of
$H_{({7\over10})}$. In fact the supercharges
$$Q = \int [Gdz + 4\lambda\bar{\psi}d\bar{z}];\quad
\bar Q = \int [\bar{G}d\bar{z} + 4\lambda\psi{dz}]$$
are fractional spin integrals of motion in the perturbed theory [14].
They were used in [14] to construct the factorizable scattering S-matrix
given below.

Along the line AT, the vacuum is three-fold degenerate, thus the particle
spectrum of (2) with $\lambda<0$ must contain ``kinks'' separating
the domains of different vacua. As shown in Fig.4a, if we label the
vacua as {$-1, 0, 1$}, then there are four sorts
of kinks of mass $m$: $K_{0,+1}, K_{0,-1}, K_{1,0}, K_{-1,0}$. An asymptotic
N-kink scattering state, say an in-state
$$|K_{\sigma_0,\sigma_1}(\theta_1)K_{\sigma_1,\sigma_2}(\theta_2)
K_{\sigma_2,\sigma_3}(\theta_3)...K_{\sigma_{N-1},\sigma_N}(\theta_N)>_{in}
\eqno(3)$$
where $\theta_i$ is the rapidity of the $i$-th kink, can be thought of
as the sequence of $N+1$ domains of vacua $\sigma_0,\sigma_1,...,\sigma_N$
placed along the x-axis, with the kink $K_{\sigma_i,\sigma_{i+1}}$
separating two neighboring domains $\sigma_i$ and $\sigma_{i+1}$. The
neighboring vacua must satisfy the ``admissibility condition''
$|\sigma_i-\sigma_{i+1}|=1$. Thus in (3) the vacua $+1$ and $-1$ must be
separated by the vacuum $0$. The out-state of N-kink scattering is related to
the in-stated (3) by the factorizable S-matrix built up from the two kink
S-matrix. This S-matrix consists of four different amplitudes (shown in
Fig.4b)
$A_0(\theta), A_1(\theta), B_0(\theta), B_1(\theta)$, defined as
$$|K_{0,s}(\theta_1)K_{s,0}(\theta_2)>_{in} =
A_0(\theta_{12})|K_{0,s}(\theta_2)K_{s,0}(\theta_1)>_{out} +
A_1(\theta_{12})|K_{0,-s}(\theta_2)K_{-s,0}(\theta_1)>_{out}; \eqno(4a)$$
$$|K_{s,0}(\theta_1)K_{0,s}(\theta_2)>_{in} =
B_0(\theta_{12})|K_{s,0}(\theta_2)K_{0,s}(\theta_1)>_{out}; \eqno(4b)$$
$$|K_{s,0}(\theta_1)K_{0,-s}(\theta_2)>_{in} =
B_1(\theta_{12})|K_{s,0}(\theta_2)K_{0,-s}(\theta_1)>_{out}; \eqno(4c)$$
where $s = +1$ or $-1$, and $\theta_{12} = \theta_1 - \theta_2$.
These amplitudes are given in [14] as
$$A_0 = cosh({\theta\over4})A(\theta);
\quad A_1 = -isinh({\theta\over4})A(\theta);$$
$$B_0 = cosh({\theta\over{4}} - {{i\pi}\over{4}})B(\theta); \quad
B_1 = cosh({\theta\over{4}} + {{i\pi}\over{4}})B(\theta). \eqno(5)$$
The ``minimal'' expressions for $A(\theta)$ and $B(\theta)$ are
$$A(\theta) = exp(\gamma\theta)S(\theta); \quad
B(\theta) = \sqrt{2}exp(-\gamma\theta)S(\theta), \eqno(6)$$
where $exp(2\pi{i}\gamma) = 2$ and
$$S(\theta) = \prod_{k=1}^{\infty}{{\Gamma(k - {\theta\over{2\pi i}})
\Gamma(k - {1\over 2} + {\theta\over{2\pi i}})}\over
{\Gamma(k + {\theta\over{2\pi i}})
\Gamma(k + {1\over 2} - {\theta\over{2\pi i}})}}. \eqno(7)$$
Note that these amplitudes do not exhibit any poles in
the ``physical strip'', hence there are no bounded state in the bulk
scattering theory.
\hbox{}\hfil\break

{\bf 4. The General Boundary S-Matrix}
\hbox{}\hfil\break

In this section we consider the perturbed Tricritical Ising model in the
semi-infinite plane $\{{(x,y):x\le0; -\infty<y<\infty}\}$. The general symbolic
action with general boundary condition has the form
$$H_{\lambda,\Phi_B} = H_{{7\over10}+CBC} + \lambda\int_{-\infty}^{\infty}
{dy}\int_{-\infty}^{0}{\Phi_{3\over5}(x,y)dx} +
\int_{-\infty}^{\infty}{\Phi_B(y)dy}, \eqno(8).$$
where $H_{{7\over10}+CBC}$ is the CFT action on the half-space with Conformal
Boundary Conditions (CBC) on the boundary $x=0$. Here we assume a choice for
the boundary operator $\Phi_B$ can be made such that the integrability
of the bulk theory is preserved.
Therefore the field theory (8) can be described by a
factorizable scattering theory of kinks with the boundary. Since the bulk
theory contains three degenerate vacua, we can expect the boundary to exist
in one or more of these vacuum states. In the general scenario, the boundary
states need not be degenerate, and some of these states can appear as boundary
bound states in the scattering process [10]. If some of the boundary
vacua are degenerate, then the scattering of kinks with the boundary can change
its state.

Let us denote the ``boundary creating operator'' by $B_a$ where
$a\in\{-1, 0, 1\}$ labels the boundary vacuum state. The Fock space of the
boundary scattering theory consists of the scattering in-states
$$|K_{\sigma_1,\sigma_2}(\theta_1)K_{\sigma_2,\sigma_3}(\theta_2)...
K_{\sigma_N,a}B_a>_{in} \eqno(9)$$
with $\theta_1>\theta_2>...>\theta_N>0$. The out-state of the scattering
is again related to the in-state (9) by a S-matrix built up from the
two-kink S-matrix (4) and the one-kink boundary scattering S-matrix.
The boundary S-matrix comprised of six elementary amplitudes defined as
$$|K_{+1,0}(\theta)B_0>_{in} = R_+(\theta)|K_{+1,0}(-\theta)B_0>_{out};
\eqno(10a)$$
$$|K_{-1,0}(\theta)B_0>_{in} = {R}_-(\theta)|K_{-1,0}(-\theta)B_0>_{out};
\eqno(10b)$$
$$|K_{0,+1}(\theta)B_{+1}>_{in} = P_+(\theta)|K_{0,+1}(-\theta)B_{+1}>_{out}
+ V_+(\theta)|K_{0,-1}(-\theta)B_{-1}>_{out}; \eqno(10c)$$
$$|K_{0,-1}(\theta)B_{-1}>_{in} = {P}_-(\theta)|K_{0,-1}(-\theta)B_{-1}>_{out}
+ V_-(\theta)|K_{0,+1}(-\theta)B_{+1}>_{out}, \eqno(10d)$$
and the associated space-time diagrams are illustrated in Fig.5.
Note that in the general case we do not assume a priori that the boundary will
respect the spin-reversal symmetry ($\sigma \to -\sigma$) of the bulk theory.

Factorizability of the scattering theory (8) means that the boundary
amplitudes (10) have to satisfy the Boundary Yang-Baxter equation
[3, 10]
(shown in Fig.6) and results in the following set of functional equations:
$$R_+A_0R_+A_1 + R_+A_1{R}_-A_0 =
{R}_-A_1R_+A_0 + {R}_-A_0{R}_-A_1; \eqno(11a)$$
$$P_+B_0V_+B_1 + V_+B_1{P}_-B_1 =
V_+B_0P_+B_0 + {P}_-B_1V_+B_0; \eqno(11b)$$
$$P_+B_1V_+B_0 + V_+B_0{P}_-B_0 =
V_+B_1P_+B_1 + {P}_-B_0V_+B_1; \eqno(11c)$$
$$P_+B_1P_+B_1 + V_+B_0V_-B_1 =
P_+B_1P_+B_1 + V_-B_0V_+B_1; \eqno(11d)$$
$$P_+B_0P_+B_0 + V_+B_1V_-B_0 =
P_+B_0P_+B_0 + V_-B_1V_+B_0, \eqno(11e)$$
(the other equations are obtained by interchanging $+ \leftrightarrow -$
in (11))
where each term in the equations has the arguments
$R_i(\theta_1)S_j(\theta_2+\theta_1)R_k(\theta_2)S_l(\theta_2-\theta_1)$.
Notice that the equation for $R_+$ and ${R}_-$ decouples from the
equations for the other amplitudes. This is to be expected since the
scattering associated with $R_+$ and ${R}_-$ does not involve changes in the
boundary state. However if the amplitudes (11) exhibit ``boundary bound
states'', then the ``boundary bound-state bootstrap equation'' will mix
the various amplitudes. This will be discussed in Section 6.
Solution to these equations can be obtained by elementary methods and
has the form
$$R_+(\theta) = (1+Ash{\theta\over2})M(\theta); \quad
{R}_-(\theta) = (1-Ash{\theta\over2})M(\theta); \eqno(12a)$$
$$P_+(\theta) = (X+Ysh{\theta})N(\theta); \quad
{P}_-(\theta) = (X-Ysh{\theta})N(\theta); \eqno(12b)$$
$$V_+(\theta) = k_+sh{\theta\over2}N(\theta); \quad
V_-(\theta) = k_-sh{\theta\over2}N(\theta), \eqno(12c)$$
where $A, X, Y, k_+$ and $k_-$ are constants depending on the boundary
condition. It is worth noting that $R_+$ and ${R}_-$ contains
one free parameter $A$, while the other amplitudes involve three independent
parameters (we can always absorb a constant into $N(\theta)$ when needed).
The normalization functions $N(\theta)$ and $M(\theta)$ can
be determined by the boundary unitarity and cross-unitarity constraint
for the boundary S-matrix. For this model, the unitarity conditions assume
the form
$$R_+(\theta)R_+(-\theta) = 1; \eqno(13a)$$
$$P_+(\theta)V_+(-\theta) + V_+(\theta){P}_-(-\theta) = 0; \eqno(13b)$$
$$P_+(\theta)P_+(-\theta) + V_+(\theta)V_-(-\theta) = 1; \eqno(13c)$$
and other equations with the substitution $+ \leftrightarrow -$ in
(13).

To obtain the cross-unitarity condition, we have to analytically continue
the boundary S-matrix (10) into the cross channel [10], resulting in the
following equations
$$R_+({{\pi{i}}\over2}-\theta) = A_0(2\theta)R_+({{\pi{i}}\over2}+\theta)
+ A_1(2\theta)R_-({{\pi{i}}\over2}+\theta); \eqno(14a)$$
$$P_+({{\pi{i}}\over2}-\theta) = B_0(2\theta)P_+({{\pi{i}}\over2}+\theta);
\eqno(14b)$$
$$V_+({{\pi{i}}\over2}-\theta) = B_1(2\theta)V_+({{\pi{i}}\over2}+\theta),
\eqno(14c)$$
and similar equations under the interchange $+ \leftrightarrow -$
in (14). The unitarity conditions (13) and (14) reduce to
$$M(\theta)M(-\theta) = {1 \over {1 - A^2sh^2{\theta\over2}}}; \eqno(15a)$$
$$M({{\pi{i}}\over2}-\theta) = e^{4\gamma\theta}B_0(2\theta)
M({{\pi{i}}\over2}+\theta); \eqno(15b)$$
$$N(\theta)N(-\theta) = {1 \over {X^2 - Y^2sh^2{\theta} -
k_+k_-sh^2{\theta\over2}}}; \eqno(15c)$$
$$N({{\pi{i}}\over2}-\theta) = B_0(2\theta)N({{\pi{i}}\over2}+\theta),
\eqno(15d)$$
for the normalization functions $N(\theta)$ and $M(\theta)$.
The solution to these equations requires analytic information about the
scattering amplitudes (12) which are governed by the boundary condition. More
specifically we need to determine the existence, if any, of ``physical poles''
(ie poles in the physical strip) in these amplitudes. To this end we will
describe in the next sections the various conformal boundary conditions in
this model and their associated scattering S-matrices.
\hbox{}\hfil\break

{\bf 5. Conformal Boundary Conditions of $c={7\over10}$ CFT}
\hbox{}\hfil\break

A vareity of boundary conditions are possible for a given bulk theory,
and they can be catagolrized into universality classes much like the case
of the bulk theory.
Each universality class of boundary conditions are characterized
by a boundary fixed point
[28] which is invariant under conformal transformations. Various
conformal fixed points can be connected by renormalization group trajectories.
The conformal boundary conditions for $c={7\over10}$ may be classified
following
the work in [2]. Conformal invariance of the boundary is equivalent to the
requirement that the two components $T$ and $\bar T$ of the stress tensor be
equal on the boundary. If we denote a conformal boundary state by $|a\r$, then
it must be annihilated by $(L_n-\bar L_n)$
for each $n$.
Solutions to this constraint were given in [29], and they are
linear combinations of the states
$$|j> = \sum_{N}|j,N>\otimes\bar{|j,N>}, \eqno(16)$$
where $j$ labels a highest weight representation of the algebra of the $L_n$,
and $|j,N>$ is an orthonormal basis in this representation space.
Similarly, $\bar{|j,N>}$ forms an orthonormal basis
in the $j$ representation of the algebra of $\bar{L}_n$.
\footnote{$^2$}{Alternatively, we could try to classified the boundary states
using the super currents $G$ and $\b{G}$ of the bulk theory [2].
The resultant
states will carry highest weight representations of the superconformal
symmetry. We will not pursue this line of analysis here.}
The boundary states corresponding to ``physical'' boundary conditions can
be constructed following the procedure in [2] and have the form

$|\tilde{0}\r = C[|0\r + \eta|{1\over10}\r + \eta|{3\over5}\r + |{3\over2}\r
+ \root4\of{2}|{7\over16}\r + \root4\of{2}|{3\over80}\r];$

$|\tilde{1\over10}\r = C[\eta^2|0\r - \eta^{-1}|{1\over10}\r
- \eta^{-1}|{3\over5}\r + \eta^2|{3\over2}\r -
\root4\of{2}\eta^2|{7\over{16}}\r
+ \root4\of{2}\eta^{-1}|{3\over{80}}\r];$

$|\tilde{3\over5}\r = C[\eta^2|0\r - \eta^{-1}|{1\over{10}}\r
- \eta^{-1}|{3\over5}\r + \eta^2|{3\over2}\r + \root4\of{2}\eta^2|{7\over16}\r
- \root4\of{2}\eta^{-1}|{3\over80}\r];$

$|\tilde{3\over2}\r = C[|0\r + \eta|{1\over10}\r + \eta|{3\over5}\r +
|{3\over2}\r
- \root4\of{2}|{7\over16}\r - \root4\of{2}|{3\over80}\r];$

$|\tilde{7\over16}\r = \sqrt2C[|0\r - \eta|{1\over10}\r + \eta|{3\over5}\r -
|{3\over2}\r] = {1\over{\sqrt2}}[{\cal{D}}(|\tilde{0}\r +
|\tilde{3\over2}\r)];$

$|\tilde{3\over80}\r = \sqrt2C[\eta^2|0\r + \eta^{-1}|{1\over10}\r
- \eta^{-1}|{3\over5}\r - \eta^2|{3\over2}\r] =
{1\over{\sqrt2}}[{\cal{D}}(|\tilde{1\over10}\r + |\tilde{3\over5}\r)],
\hfil (17)$

\noindent where $C = \sqrt{{sin{\pi\over5}}\over{\sqrt5}}$ and
$\eta = \sqrt{{sin{{2\pi}\over5}}\over{sin{\pi\over5}}}$. Here $\cal{D}$
denotes the Kramers-Wannier duality transform. Thus we can regard the
state $|\tilde{7\over16}\r$ as the dual boundary condition
of $|\tilde{0}\r$ and $|\tilde{3\over2}\r$. Similarly $|\tilde{3\over80}\r$
is dual to $|\tilde{3\over5}\r$ and $|\tilde{1\over10}\r$.

To interpret
the boundary conditions associated with the states in (17), it will
be illustrative to refer to the $A_4$ RSOS lattice realization of this
model [21]. Consider a diagonal square lattice where the degree of
freedom at each site can take on four values $l_i = 1, 2, 3$ and $4$.
The value of $l_i$ is further constrained by the requirement
$|l_i-l_j| = 1$ for nearest neighbor sites $i$ and $j$. In addition
there is a ferromagnetic next-nearest-neighbor interaction and a
single-site interaction. Thus the lattice divides into two sublattices
I and II, where the values of $l_i$ are odd in one sublattice, and
even in the other sublattice. It is clear that there are three degenerate
ground states as shown in Fig.9. To make contact with the order
parameter $<\sigma>$ of the Ising model
with vacancies (1), we shall label the ground states in Fig.9
by $-1, 0$ and $+1$.
The Hamiltonian of the RSOS model is symmetric under
the global $Z_2$ transform $l_i \to 5 - l_i$, which corresponds to the
spin-reversal symmetry of the model (1). The three ground states will
become identical in the bulk at the tricritical point. However the nature
of the order parameter at the boundary will depend on the specific boundary
condition. As shown in [2], the state $|\tilde{\Delta}_{(1,s)}\r$ corresponds
to the
boundary condition where all boundary degrees of freedom are fixed to the
value $s$, this is illustrated in Fig.10.
For the state $|\tilde{0}\r$, the boundary degrees of freedom are fixed to
$1$, and hence the neighboring state must be in the state $2$. Thus in
the continuum limit, the order parameter at the boundary will be in the $-1$
vacuum, and we shall denote this boundary condition by $(-)$. The boundary
state $|\tilde{1\over10}\r$ corresponds to fixing the the boundary degree
of freedom to $2$, and the neighboring sites can be in either states $1$ or
$3$. In this case the order parameter at the boundary may be in the $-1$ or
$0$ vacuum. This boundary condition where the $-1$ and $0$ vacua are
degenerate will be labeled $(-0)$.
We can similarly associated $|\tilde{3\over5}\r$ with the boundary condition
$(0+)$ where the $0$ and $+1$ vacua are degenerate at the boundary. In the
same way, the boundary condition $(+)$ where the order parameter is fixed to
the $+1$ vacuum will correspond to the boundary state $|\tilde{3\over2}\r$.
One can regard $(-)$ and $(+)$ as ``fixed'' boundary conditions for the
Ising model with vacancies (1).

In the work of [30], they identified the boundary states
$|\tilde{\Delta}_{(r,1)}>$ as the boundary condition where the boundary
degrees of freedom are fixed to the value $r$, while the neighboring sites
must be in the state $r+1$ (for $1\le{r}\le{3}$). For the boundary state
$|\tilde{7\over16}\r$, this means that the boundary will be in state $2$
and the neighboring sites will be in state $3$. This of course fixes
the order parameter to be in the vacuum $0$ in the continuum limit, and
we denote this the $(0)$ boundary condition. A few comments about this
boundary condition is in order. Using the states in (17), the partition
function of the model on the cylinder with boundaries at both ends can be
computed. Let us define the modular parameter as $q \equiv e^{2\pi{i}\tau}$,
where $\tau = {{iR}\over{2L}}$ on a cylinder of length $L$ and circumference
$R$. The partition functions involving the boundary condition $(0)$ include
$$Z_{(0)(0)}(q) = \chi_{0}(q) + \chi_{3\over2}(q), \eqno(18a)$$ and
$$Z_{(-)(0)}(q) = \chi_{7\over16}(q). \eqno(18b)$$
The result (18a) was conjectured earlier in [31] for the Tricritical Ising
partition function with microscopic ``free'' boundary spins, and later
confirmed
by direct calculation [30] and numerically in [32, 33].
Furthermore (18b) was also obtained in [33] by considering the partition
function with ``mixed'' boundary condition, ie microscopic ``fixed'' and
``free'' boundary conditions.
Thus we concluded that the boundary condition $(0)$ is associated
with the lattice Tricritical Ising model where the microscopic boundary spins
are not constrained. Such lattice boundary conditions are often referred to
in the literature as the ``free'' boundary condition, even in the conformal
field theory.

Lastly we consider the conformal boundary state $|\tilde{3\over80}\r$ which
has not previously been identified. Note that it is dual to the boundary
conditions $(-0)$ and $(0+)$, and we conjecture that it corresponds to the
case where the boundary can exist in all three vacua. In other words, like
in the bulk theory, the three vacua are also degenerate on the boundary. We
shall call such boundary condition ``degenerate'' and label it as $(d)$.
Since a great deal of ``fine-tuning'' of the boundary parameters are required
to achieve three-fold degeneracy of the vacua, we expect the boundary
condition $(d)$ to be unstable under perturbation by relevant boundary
operators. This boundary condition carries the $3\over{80}$ representation of
the Virasoro algebra, so the boundary operators that can appear along such
a boundary is given by the Operator Product Expansion of the primary field
$\Phi_{({3\over{80}})}$ with itself [18]. From the operator product
$$\Phi_{({3\over{80}})}\Phi_{({3\over{80}})} = [\Phi_{(0)}] +
[\Phi_{({1\over{10}})}] + [\Phi_{({3\over5})}] + [\Phi_{({3\over2})}],$$
we see that two relevant boundary fields $\psi_{({3\over5})}$ and
$\psi_{({1\over{10}})}$ can appear for this boundary condition. Perturbing by
these relevant operators could generate renormalization group flow from the
boundary condition $(d)$ down to a more ``stable'' boundary fixed point.
To lend support for this speculation, we computed the boundary ground state
degeneracy (the so-called ``$g$-factor'' [34])
from the boundary states (17) and
they are in order of decreasing magnitude
$$g_{(d)} = \sqrt{2}\eta^2C; \quad g_{(-0)} = g_{(0+)} = \eta^2C,$$
$$g_{(0)} = \sqrt{2}C; \quad \hbox{and} \quad g_{(-)} = g_{(+)} = C.$$
The value of $g$ should decrease along renormalization group trajectories,
and in this respect, $(d)$ would be the most ``unstable'' conformal boundary
condition. Similar comments can be made about the two-fold degenerate
boundary conditions $(-0)$ and $(0+)$ with the next-largest value of $g$.
One can repeat the analysis for the associated boundary operator for these
boundary conditions, and we found that they both admit the relevant boundary
operator $\psi_{({3\over5})}$. Thus we expect that the perturbation of
$(-0)$ and $(0+)$ by this operator would
generate RG flow down to other conformal boundary conditions. This is studied
in the next section.
\hbox{}\hfil\break

{\bf 6. Boundary S-matrix for the Conformal Boundary Conditions}
\hbox{}\hfil\break

Boundary conditions can in general be changed by altering the coupling
constants of model on the boundary. A simple example would be the
introduction of a ``magnetic field'' which favors a particular direction of
lattice spins. In the continuum limit, this magnetic field would shift the
specific free energy of one of the vacua and destroy the ground state
degeneracy.
Let us consider the massive Tricritical Ising model with the conformal
boundary condition $(-0)$ and perturbed by the relevant boundary operator
$\psi_{({3\over5})}$. The symbolic action for the corresponding field theory
is (8) with
$$H_{{7\over{10}}+CBC} = H_{{7\over{10}}+(-0)};
\quad \Phi_B = h\psi_{({3\over5})}, \eqno(19)$$
and $h$ is some boundary coupling constant with dimension
$[length]^{-{2\over5}}$. Since the boundary operator
has the same dimension as the bulk perturbing operator $\Phi_{({3\over5})}$,
this boundary perturbation is integrable [10]. The boundary field
$\psi_{({3\over5})}$ couples to the boundary spins for the boundary condition
$(-0)$ and would alter the specific free energy of the $-1$ vacuum. Thus for
$h\ne{0}$, the ground state will no longer be degenerate.

To be more specific, let us treat the case of a RG flow from $(-0)$ down
to the ``fixed'' boundary condition $(-)$. For $h>{0}$ we expect the
vacuum $-1$ to be the ground state of the boundary, while the vacuum $0$
is an excited state. The associated boundary scattering S-matrix will involve
the amplitude ${P}_-(\theta)$, and since the boundary cannot exist in
the vacuum $+1$, we have $k_-=k_+=0$ in this case. The Yang-Baxter equation
(11) becomes an identity for this boundary condition, and the amplitude
${P}_-(\theta)$ will be determined by the unitarity conditions (13)
and (14). The solution can be written in the form
$${P}_-(\theta) = P_{\xi}^{CDD}(\theta)P_{min}(\theta), \eqno(20)$$
where $P_{min}(\theta)$ is the minimal solution of the equations
$$P_{min}(\theta)P_{min}(-\theta) = 1; \eqno(21a)$$
$$P_{min}({{i\pi}\over2}-\h) = B_0(2\h)P_{min}({{i\pi}\over2}+\h),
\eqno(21b)$$
with no poles in the physical strip, and has the expression
$$P_{min}(\h) =
\prod_{k=1}^{\infty}{{{\Gamma(k - {\theta\over{2\pi i}})}^2
\Gamma(k - {1\over 4} + {\theta\over{2\pi i}})
\Gamma(k + {1\over 4} + {\theta\over{2\pi i}}))}\over
{{\Gamma(k + {\theta\over{2\pi i}})}^2
\Gamma(k + {1\over 4} - {\theta\over{2\pi i}})
\Gamma(k - {1\over 4} - {\theta\over{2\pi i}})}}\eqno(22).$$
The other factor
$$P_{\xi}^{CDD}(\h) = {{sin{\xi}-ish{\h}}\over{sin{\xi}+ish{\h}}}, \eqno(23)$$
is a CDD factor [35] which exhibits a pole at $\h = i\xi$. When this pole
lies inside the ``physical strip'', $0\le{\xi}\le{{\pi}\over2}$, it could be
interpreted as a ``boundary bound state'' as shown in Fig.11. If we denote
$e_a$ to be the specific energy (per unit boundary length) of the boundary
state $|B_a>$, then
$$e_0 - e_{-1} = mcos{\xi}, \eqno(24)$$
defines the binding energy of the bound state.
Clearly the parameter $\xi$ is related to the boundary coupling constant $h$,
and we can analyzes the effect of the RG flow by varying the value of $\xi$.
Furthermore we can consider
scattering of this bound state with other kinks by the ``boundary bound
state bootstrap equation'' [10]
\footnote{$^3$}{Similar ``fusion'' procedure was used in [36] to derive
the boundary S-matrix for the Sine-Gordon model at the supersymmetric
points}
$$R_+(\h) = B_1(\h-i\xi)P_-(\h)B_1(\h+i\xi); \eqno(25a)$$
and
$$R_-(\h) = B_0(\h-i\xi)P_-(\h)B_0(\h+i\xi), \eqno(25b)$$
as illustrated in Fig.12. This gives two more scattering amplitudes which
can occur when the boundary is in the vacuum state $0$. Equation (25) can
be simplified to a form resembling the general solution (12):
$$R_+(\h) = {1\over2}(cos{\xi\over2} + ish{\h\over2})
B(\h-i\xi)B(\h+i\xi)P_{\xi}^{CDD}(\h)P_{min}(\h); \eqno(26a)$$
and
$${R}_-(\h) = {1\over2}(cos{\xi\over2} - ish{\h\over2})
B(\h-i\xi)B(\h+i\xi)P_{\xi}^{CDD}(\h)P_{min}(\h). \eqno(26b)$$
For the boundary condition $(-0)$ the vacua $-1$ and $0$ are degenerate
on the boundary and we expect the boundary to admit a bound
state at $\h = {{i\pi}\over2}$. Thus the amplitudes (20) and (26)
with $\xi = {\pi\over2}$ constitute the boundary S-matrix for this unperturbed
boundary condition $(-0)$.
When we ``turn on'' the  boundary perturbation ($h > 0$),
$(e_{0} - e_{-1}) > 0$ and the amplitude ${P}_-(\h)$ will exhibit a pole in
the physical strip at $\h = i\xi$ for $0 < \xi < {\pi\over2}$.
One can think of the bound state as formally ``gluing'' a kink $K_{0,-1}$
of rapidity $i\xi$ (interpreted as a fluctuating domain wall) to the
boundary $|B_{-1}>$. As we increase the perturbation,
$\xi \to 0$, the bound state becomes weakly bound
and the domain wall develops large fluctuations which propagate well into
the bulk. At $\xi = 0$, the domain wall has zero rapidity and is no longer
bounded to the boundary.

For $-{{\pi}\over2} < \xi \le 0$, the pole leaves the physical
strip. In this regime $(e_{0} - e_{-1}) > m$, so there is
no ``boundary bound state'' and $-1$ remains as the only
stable ground state of the boundary. At $\xi = -{{\pi}\over2}$, the poles
and zeros of $P_{\xi}^{CDD}(\h)$ become two-fold degenerate. As we increase $h$
further, $\xi$ develops an imaginary component: $\xi = -{{\pi}\over2} +
i\vartheta$ and the amplitude ${P}_-(\h)$ exhibits two poles at
$\h = -{{i\pi}\over2} \pm \vartheta$. These poles depart to infinity
when $h$ gets increasingly large. In this limit, the energy difference
between the two vacua $(e_{0} - e_{-1})$ becomes infinitely large
and we expect the resultant boundary condition to be the ``fixed'' case $(-)$.
Thus the amplitude ${P}_-$ will become the boundary S-matrix for the $(-)$
boundary condition in this limit, and we found
$$R_{(-)}(\theta) = P_{min}(\theta). \eqno(27)$$
It is not surprising that this ``flow'' from a two-fold degenerate boundary
condition down to a unique ground state resembles closely the ``flow'' from
free to fixed boundary condition in the Ising model [10].

It is interesting to consider the perturbed action (8) for $h < 0$. In this
case we expect the perturbation to ``raise'' the specific boundary
energy of the $-1$ vacuum, and the vacuum $0$ would now become the unique
ground state. This situation corresponds to the analytic continuation of
$\xi$ above the degenerate point ${\pi}\over2$. In the regime
${\pi\over2} < \xi < \pi$, the difference in boundary energies become
$$(e_{-1} - e_{0}) = mcos{\zeta} > 0, \eqno(28)$$
where $\zeta = \pi - \xi$. Hence we can regard the $-1$ vacuum as a boundary
excited state and it could appear as a boundary bound state in the scattering
S-matrix. Indeed one can check that the amplitude ${R}_-(\theta)$ now
possesses a  bound state pole in the physical strip at $\h = i\zeta$. The
other amplitude $R_+(\theta)$ does not have this pole, consistent with the
fact that the boundary cannot be in the vacuum state $+1$. The analysis in
this regime mirrors that of the domain $0 \le \xi \le {\pi\over2}$, except
that the role of $-1$ and $0$ are interchanged. At $\xi = \pi$, the bound
state becomes unbounded and $0$ emerges as the only stable boundary ground
state. As we decrease the value of $h$ further such that
$\xi > {\pi}$, $R_+(\h)$ and ${R}_-(\h)$ are the only physical
boundary scattering amplitudes and there is no more bound state pole.
When the RG flow reaches the point for which $\xi = 2\pi$, the poles
and zeros in the factor $(cos{\xi\over2} \pm ish{\h\over2})B(\h+\xi)
B(\h-\xi)P_{\xi}^{CDD}(\h)$ (common to both $R_+$ and ${R}_-$)
become two-fold degenerate. Further perturbation would introduce an imaginary
component to $\xi$: $\xi = 2\pi + i\varrho$. The amplitudes $R_+(\theta)$
and ${R}_-(\h)$ would correspondingly exhibit two poles at $\h = 2\pi{i} \pm
\varrho$. These poles will separate and depart to infinity when $h$ approach
negative infinity. Under this RG flow, we expect the boundary specific energy
$e_{-1}$ to become increasingly large, and thus the boundary will be ``fixed''
to the stable ground state $0$. This is the $(0)$ boundary condition and in
this limit $R_+$ and ${R}_-$ will both be reduced to the boundary S-matrix
for $(0)$:
$$R_{(0)}(\h) = e^{-2\gamma{\h}}P_{min}(\h). \eqno(29)$$
The RG flows from $(-0)$ boundary condition down to $(-)$ and $(0)$ are shown
schematically in Fig.13. Of course we could also consider the perturbation
of the $(0+)$ boundary condition by $\psi_{3\over5}$
and we expect it to generate the same flow
pattern and conformal boundary S-matrices. It is not easy to relate the
perturbation coupling $h$ to the S-matrix parameter $\xi$. However, based
upon the pattern of flow of the scattering amplitudes, we conjecture
the relation
in this case to be
$$h \sim m^{2\over5}[cos({{2\xi}\over5} + {{\pi}\over5}) - cos{{2\pi}\over5}].
\eqno(30)$$

As shown in Fig.13, by applying the duality transformation to the RG
flows from $(-0)$, we obtained the RG flows from the $(d)$ boundary condition
down to either $(0)$ or a superposition of $(-)$ and $(+)$. Note that the
field $\psi_{({3\over5})}$ does not change sign under duality. The symbolic
action of this dual theory is (8), where $H_{{7\over10}+CBC} =
H_{{7\over10}+(d)}$ is the CFT action with conformal boundary condition $(d)$.
For this new RG flow, the point $h = 0$ corresponds to the
boundary condition $(d)$ where all three vacua are degenerate.
This boundary condition respects the full symmetry of the bulk theory, so
the associated boundary S-matrix should be invariant under spin-reversal.
Thus we have $A = Y = 0$ and $k_+ = k_-$ in the general solution (12),
and denotes
$$R_+ = {R}_- = R; \quad P_+ = {P}_- = P \quad \hbox{and}
\quad V_+ = V_- = V.$$
When this conformal boundary condition is perturbed
with $h > 0$, the boundary energies $e_+$ and $e_-$ are both higher than
$e_0$, and hence the vacua $-1$ and $+1$ are the excited states of the
boundary. Here we shall assume the perturbation does not destroy the $Z_2$
spin-reversal symmetry of $(d)$. The amplitude $R(\h)$ can be obtained
by solving the unitarity constraints (13) and (14). It must exhibit a
bound state pole in the physical strip, say at $i\u$, corresponding to these
excited boundary states, and has the form
$$R(\h) = e^{-2\gamma\h}P_{\u}^{CDD}(\h)P_{min}(\h); \eqno(31)$$
where $P_{\u}^{CDD}(\h)$ is the CDD factor (23).
The other amplitudes are again
related to $R(\h)$ by the ``boundary bound state bootstrap equation''
$$P(\h) = [A_0(\h-\u)A_0(\h+\u) + A_1(\h-\u)A_1(\h+\u)]R(\h), \eqno(32a)$$
and
$$V(\h) = [A_0(\h-\u)A_1(\h+\u) + A_1(\h-\u)A_0(\h+\u)]R(\h). \eqno(32b)$$
The expressions for $P$ and $V$ can be simplified to
$$P(\h) =
ch{\u\over2}e^{-2\gamma\h}A(\h-\u)A(\h+\u)P_{\u}^{CDD}(\h)P_{min}(\h);
\eqno(33a)$$
$$V(\h) =
-ish{\h\over2}e^{-2\gamma\h}A(\h-\u)A(\h+\u)P_{\u}^{CDD}(\h)P_{min}(\h),
\eqno(33b)$$
which are the general solution (12) with $X = ch{\u\over2}$ and $k_+ = k_- =
-i$.

For the boundary condition $(d)$, the three vacua are degenerate and all
scattering amplitudes should possess a pole at $\h = {{i\pi}\over2}$. The
associated boundary S-matrix is given by (31) and (33) with
$\u = {\pi\over2}$. Perturbing this boundary condition in (8) with $h > 0$
would lower $e_0$, while the other vacua $-1$ and $+1$ will appear as bound
states in the amplitude $R$ at $\h = i\u$. In the domain $-{{\pi}\over2} < \u
< 0$, there is no bound state, and $R$ remains as the only physical amplitude.
Under further RG flow, $\u$ develops an imaginary component ($\u = -{\pi\over2}
+i\upsilon_0$) and travels
to infinity.
One can check that in this limit, $R$ reduces to the amplitude $R_{(0)}$ for
the boundary condition $(0)$.

Perturbing $(d)$ with $h < 0$ raises the energy $e_0$ above $e_+$ and $e_-$.
For
${\pi\over2} < \u < {\pi}$ the vacuum $0$ enters as the bound state in $P$
and $V$. In the regime ${\pi} < \u < 2\pi$, the bound state leaves
the physical strip, with $+1$ and $-1$ remaining as the only stable boundary
ground states. At $\u = 2\pi$, the poles and zeros of the
$\u$-dependent factors of both $P$ and $V$
become two-fold degenerate. Further RG flow
would introduce an ever increasing imaginary part to $\u$, ie $\u = 2\pi +
i\upsilon'$, and drives it to infinity. In this limit, the amplitude $V$
approaches zero, while $P$ flows to the ``fixed'' boundary condition
amplitude (27). This means that the scattering theory decouples into two
equivalent sectors, and the boundary becomes a superposition of the pure
boundary states
$|B_{+1}>$ and $|B_{-1}>$. Since this RG flow pattern is the same as the
flow from the $(-0)$ boundary condition, we expect the parameter $\u$ to
be related to the coupling $h$ in the same manner as (30).
\hbox{}\hfil\break

{\bf 7. Discussion}
\hbox{}\hfil\break

In this work we computed the boundary S-matrix associated with the conformal
boundary conditions of the Tricritical Ising model. The interpolating S-matrix
which describe the renormalization group flow between various boundary
conditions was also found. However we have not yet found an integrable
boundary perturbation which give rise to the general S-matrix (12). It
would be interesting to determine this more general perturbation. Since the
general solution (12) contains more than one boundary parameter, it is
conceivable that the associated perturbation involves more than one boundary
operator, and some of them would break the spin-reversal symmetry.
Another interesting follow-up
to this work would be to consider the massless limits of the S-matrix. Since
the interpolating S-matrix contains a boundary parameter, its massless limit
might be non-trivial. This information would be relevant to the interpretation
of the $c = {7\over10}$ conformal field theory as a massless scattering theory
[16, 37].

Finally, we have not discussed in this work the possibility of a supersymmetric
boundary condition [36]. Since the degenerate boundary condition $(d)$
respects the full symmetry of the bulk theory, it is plausible that it might
also preserve supersymmetry. One would like to check if any other conformal
boundary states (or a linear combination of them) could also be supersymmetric.
Furthermore, if some of the conformal boundary states do possess supersymmetry,
it would be interesting to see if it could survive under perturbation, for
example, by the boundary field $\psi_{({3\over5})}$. At the level of the
boundary scattering theory, one could check supersymmetry by determining
the appropriate linear combination of supercharges $Q$ and $\b{Q}$ which
survive as an integral of motion. This combination should commute with
the supersymmetric boundary S-matrix, and could possibly involve the
topological
charge of the massive theory [36].
\hbox{}\hfil\break

{\bf Acknowledgement}
\hbox{}\hfil\break
I would like to thank Prof. A.B.Zamolodchikov for suggesting this interesting
problem and many useful discussions.
I am grateful to Prof. R.Nepoemchie for reading the manuscript and pointing
out some mistakes. I would also like to thank E.Lo for help
in producing some of the figures.
{\bf References}
\hbox{}\hfil\break

[1]For a review, see A.W.W.Ludwig. Int. J. Mod. Phys. B8 (1994) 347.

[2] J.L.Cardy. Nucl. Phys. B324 (1989) 581.

[3] I. Cherednik. Theor. Math. Phys. 61, 35 (1984) 977.

[4] E.Sklyanin. J. Phys. A21 (1988) 2375.

[5] L.Mezincescu and R.I.Nepomechie. J. Phys. A24 (1991) L17.

[6] L.Mezincescu and R.I.Nepomechie and V.Rittenberg. Phys. Lett. 147A (1990)
70.

L.Mezincescu and R.I.Nepomechie. J. Phys. A25 (1992) 2533.

[7] L.Mezincescu and R.I.Nepomechie. Nucl. Phys. B372 (1992) 597.

[8] C.Destri and H.J.de Vega. Nucl. Phys. B374 (1992) 692; Nucl. Phys. B385
(1992) 361.

[9] C.M.Yung and M.T.Batchelor. Nucl. Phys. B435 (1995) 430.

[10] S.Ghoshal and A.B.Zamolodchikov. Int. J. Mod. Phys. A9 (1994) 3841.

[11] P.Fendley and H.Saleur. ``Exact theory of polymer adsorption in
analogy with the Kondo problem''. Preprint USC-94-006, cond-mat/9403095.

[12] P.Fendley, A.Ludwig and H.Saleur. ``Exact conductance through point
contacts in the $\nu = 1/3$ Fractional Quantum Hall Effect''.
Preprint UC-94-12, PUPT-94-1491, cond-mat/9408068.

[13] P.Fendley, F.Lesage and H.Saleur. ``Solving 1-D plasmas and 2-D
boundary problems using Jack polynomials and functional relations''. Preprint
hepth/9409176.

[14] A.B.Zamolodchikov. ``Fractional Spin Integrals of Motion in 2D
Conformal Field Theory'', in ``Fields, Strings and Quantum Gravity'',
H.Guo, Z.Qiu and H.Tye eds., Gordon and Breach, 1989.

[15] D.A.Kastor, E.J.Martinec and S.Shenker. Nucl. Phys. B316 (1989) 590.

[16] Al.B.Zamolodchikov. Nucl. Phys. B358 (1991) 524.

[17] J.D.Lawrie and S.Sarbach, in ``Phase transitions
and critical phenomena'', vol. 9, C.Domb and J.Lebowitz eds., Academic Press,
1984.

[18] D.Friedan, Z.Qui and S.Shenker. Phys. Lett. 151B (1985) 37.
M.A.Bershadski, V.G.Knizhnik, M.G.Teitelman. Phys. Lett. 147B (1985)
217.

[19] M.J.Tejwani, O.Ferreira and O.E.Vilches. Phys. Rev. Lett. 44 (1980)
152.
W.Kinzel, M.Schick and A.N.Berker, in ``Ordering in two dimensions'',
Sinha ed., North-Holland, 1980.

[20] Y.Gefen, Y.Imry and D.Mukamel. Phys. Rev. B23 (1981) 6099.
F.C.Alcaraz, J.R.Drugowich de Felicio, R.Koberle and J.F.Stilck.
Phys. Rev. B32 (1985) 7469.

[21] G.E.Andrews, R.J.Baxter and P.J.Forrester. J. Stat. Phys. 35 (1984) 93.

[22] M.J.Blume, V.J.Emery and R.B.Griffiths. Phys. Rev. A4 (1971) 1071.

B.Nienhuis, A.N.Berker, E.K.Riedel and M.Schick. Phys. Rev. Lett. 43
(1979) 737.

[23] A.B.Zamolodchikov. Sov. J. Nucl. Phys. 44 (1986) 529.

[24] D.Friedan, Z.Qui and S.Shenker. Phys. Rev. Lett. 52 (1984) 1575.

[25] L.Kadanoff and H.Ceva. Phys. Rev. B3 (1970) 3918.

E.Fradkin and L.Kadanoff. Nucl. Phys. B170 (1980) 1.

[26] M.Lassig, G.Mussardo and J.Cardy. Nucl. Phys. B348 (1991) 591.

[27] A.B.Zamolodchikov, in ``Advanced Studies in Pure Mathematics'',
vol. 19 (1989) 641.

[28] H.W.Diehl, in ``Phase transitions and critical phenomena'' vol. 10,
C.Domb and J.L.Lebowitz eds., Academic Press, 1986.

[29] N.Ishibashi. Mod. Phys. Lett. A4 (1989) 251.

[30] H.Saleur and M.Bauer. Nucl. Phys. B320 (1989) 591.

[31] J.L.Cardy. Nucl.Phys. B275 (1986) 200.

[32] D.B.Balbao and J.R.Drugowich de Felicio. J. Phys. A20 (1987) L207.

[33] U.Grimm and V.Rittenberg. Int. J. Mod. Phys. B4 (1990) 969.

[34] I.Affleck and W.W.Ludwig. Phys. Rev. Lett. 67 (1991) 161.

[35] L.Castillejo, R.H.Dalitz and F.J.Dyson. Phys. Rev. 101 (1956) 453.

[36] N.P.Warner. ``Supersymmetry in boundary integrable models'',
Preprint USC-95/014, hep-th/9506064.

[37] P.Fendley and H.Saleur. ``Massless integrable quantum field theories
and massless scattering in 1+1 dimensions''. Preprint USC-93-022,
hep-th/9310058.
\vfill\eject
{\bf Figure Captions}

Fig.5.1. Phase diagram of the Tricritical Ising model.

Fig.5.2. Landau-Ginzburg effective potential for the various regions
of the phase diagram.

Fig.5.3. Kac table of conformal dimensions for $c = {7\over10}$.

Fig.5.4. Space-time diagrams of the four variety of the kinks (a) and
their bulk scattering amplitudes (b).

Fig.5.5. Space-time diagrams for the General Boundary Scattering
Amplitudes.

Fig.5.6. Figure of the ``Boundary Yang-Baxter'' equation satisfied by
the boundary S-matrix. Here $a,b,c\in\{+1,-1\}$.

Fig.5.7. Space-time diagram for the Bundary Unitarity equation with
$a,d\in\{+1,-1\}$.

Fig.5.8. Figues of the Cross-Unitarity condition with $a,b\in\{+1,-1\}$.

Fig.5.9. The three ground states of the RSOS model and the
associated vacuum states in the continuum limit.

Fig.5.10. Various conformal conditions of the Tricritical Ising model in
terms of the RSOS picture.

Fig.5.11. Bound state pole exhibit by the amplitude $P_-$.

Fig.5.12. Figure of the ``Boundary Bound State Bootstrap'' equation with

$a\in\{+1,-1\}$.

Fig.5.13. Schematic diagram for the renormalization group flow from the
$(-0)$ and $(d)$ boundary conditions. Here $\cal{D}$ denotes duality
transformation.

Fig.5.14. Figure of the ``Boundary Bound State Bootstrap'' equation for
the perturbation of the degenerate boundary condition. Here
$a,b\in\{+1,-1\}$.
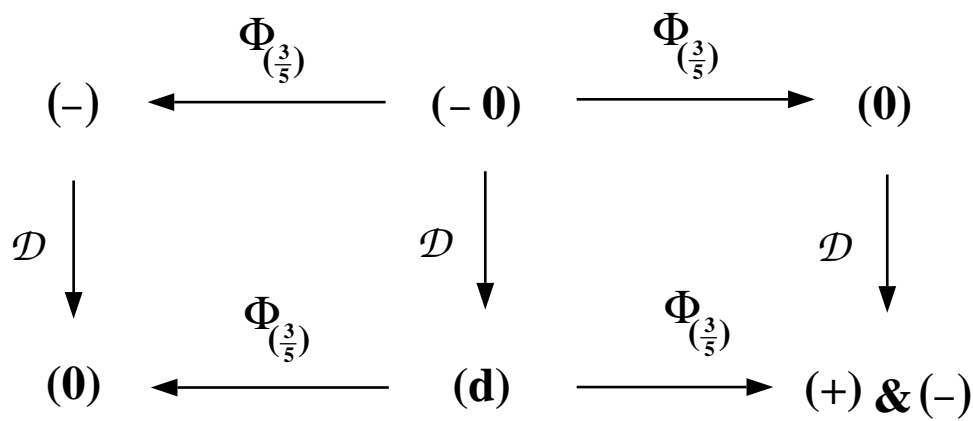
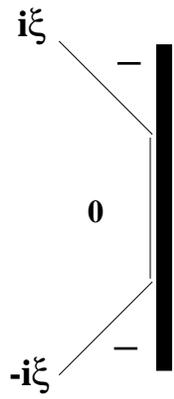
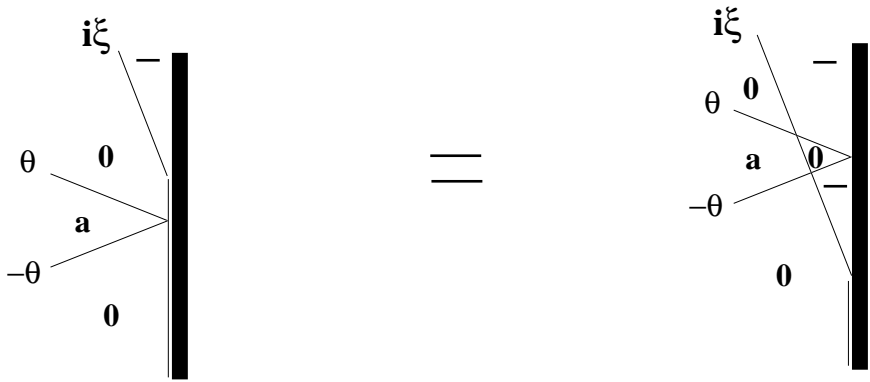
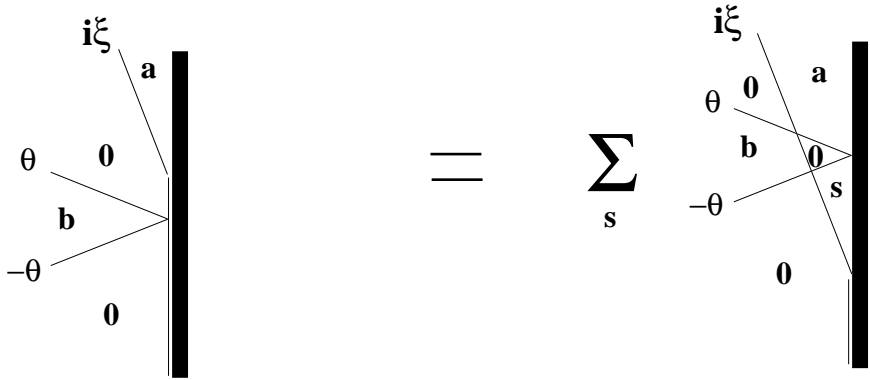
\end